\documentclass[11pt, fceqn]{article}
 \usepackage{amsfonts}
 \usepackage{flafter}
 \usepackage{amssymb,graphics,graphicx,subfigure,caption2,rotating}
 \usepackage{amsmath, latexsym}
\usepackage[amsmath,thmmarks]{ntheorem}
 \usepackage[top=3.5cm]{geometry}
\begin{document}
\date{}
\title{A bipartite separable ball and its applications}
\author{Shu-Qian Shen$^1$\thanks{E-mail: sqshen@upc.edu.cn.}, Ming Li$^1$\thanks{E-mail: liming@upc.edu.cn. }, Lei Li$^{1,2}$ \thanks{E-mail: lileiupc@126.com.}\\
{\small{\it $^{1}$College of Science, China University of Petroleum, Qingdao, 266580, P.R. China}}\\
{\small{\it $^2$Department of Mathematics, Shanghai University, Shanghai, 200444, P.R. China}}
}\maketitle
\begin{abstract}
In this paper, based on a matrix norm, we first present a ball of separable unnormalized states around the identity matrix  for the bipartite quantum system, which is larger than the separable ball in Frobenius norm. Then the proposed ball is used to get not only simple sufficient conditions for the separability of pseudopure states and the states with strong positive partial transposes, but also a separable ball centered at the identity matrix for the multipartite quantum system.
\end{abstract}

\section{Introduction}

Quantum entanglement plays a key role in the fields of quantum information processing, including quantum teleportation \cite{teleportation}, quantum computing \cite{computation}, and quantum communications \cite{communication}, etc. However, it is extremely difficult to check whether a given quantum state is entangled or separable (not entangled) \cite{NPhard}.  From \cite{Zyczkowski1998} we know that the set of separable states forms a convex body, and all states sufficiently close to the maximally mixed state are necessarily separable. Thus, it is of great interest to describe the largest balls of separable states in various norms which fit in this convex body.

In the Schatten $p$-norm with $1\le p\le \infty$, the balls of separable unnormalized states around the identity matrix were studied in \cite{Gurvits2002} for the bipartite quantum system. The authors showed that these balls are the largest ones in Schatten $p$-norms, i.e., any ball centered at the identity matrix with larger radius must contain entangled states or matrices without positive semidefiniteness. Soon after \cite{Gurvits2002}, Gurvits and Barnum \cite{Gurvits2003-2,Gurvits2005} presented the separable balls in Frobenius norm for multipartite quantum system, which have exponentially larger radius than the existing one given in \cite{Braunstein1999}. Unlike the ball for bipartite quantum system, these balls are not necessarily the largest ones. They were further improved by Hildebrand in \cite{Hildebrand2007} for the multiqubit quantum system.
These separable balls can be used to provide not only some simple sufficient separable conditions for the pesudopure states arising from the technique of liquid-state nuclear magnetic resonance (NMR), but also the tools to study the complexity of the problem about entanglement and separability; see, e.g., \cite{Gurvits2002}-\cite{Hildebrand2007}.

All the separable balls mentioned above are due to the Schatten $p$-norm with $1\le p\le \infty$, in particular, the Frobenius norm. Based on a new norm, this paper is first devoted to obtaining another separable ball for the bipartite quantum system, which is larger than the ball given in \cite{Gurvits2002} with Frobenius norm. By making use of this ball, we then derive some simple sufficient separable conditions for the pseudopure states \cite{Gurvits2002,Gurvits2003-2,pseudopure} and the states with strong positive partial transposes \cite{Chruscinski2008}-\cite{Bylicka2013}. Finally, a separable ball depending on the new norm is obtained for the multipartite quantum system. As an example, an unnormalized state is shown to be in this ball, but not in the balls given by \cite{Gurvits2003-2,Gurvits2005,Hildebrand2007}.

The remainder of the paper is organized as follows. In Section 2, we introduce the notation, preliminaries, and the  new norm. In Section 3, the separable ball for bipartite quantum system is given firstly. Secondly, this ball is used to investigate the separability of the pseudopure states and the states with strong positive partial transposes. Finally, the presented separable ball for bipartite quantum system is generalized to the multipartite quantum system. In Section 4, some concluding remarks are given.

\section{Notation and preliminaries}
We denote by $\mathbb{H}\mathbb{C}^{n\times n}$ and $\mathbb{P}\mathbb{C}^{n\times n}$ the sets of $n\times n$ Hermitian complex matrices and $n\times n$ Hermitian positive semidefinite matrices, respectively. For any $X\in \mathbb{C}^{n\times n}$, $X^\dag$ denotes the conjugate transpose of $X$, and $X$ can be decomposed into $X=H(X)+S(X)$ with $H(X)=\frac{1}{2}(X+X^\dag)$ and $S(X)=\frac{1}{2}(X-X^\dag)$. The Schatten $p$-norm of $X$ is defined as
\[
\left\|X\right\|_p=\left(\text{tr}\left((X^\dag X)^{\frac{p}{2}}\right)\right)^{\frac{1}{p}},\quad 1\le p\le \infty.
\] As is well known, $||X||_2$ and $||X||_\infty$ are the Frobenius norm and the spectral norm of $X$, respectively.

 The following definitions about cones can be found in \cite{Gurvits2003-2}. A cone is a subset of a vector space that is closed under multiplication by positive scalars. Let $C_1\subset V_1$ and $C_2\subset V_2$ be two cones. Then the linear map $\phi:V_1\rightarrow V_2$ is said to be $C_1$-to-$C_2$ positive if $\phi(C_1)\subseteq C_2$. When $C_2$ is chosen to be the set of the Hermitian positive semidefinite matrices, ``$C_1$-to-$C_2$ positive" is abbreviated to be ``$C_1$-positive". The cone $S$ is said to be generated from the set $T$ if and only if $S$ is the set of the positive linear combination of elements of $T$.

A state represented by a matrix $\rho$ (not necessarily normalized, but Hermitian positive semidefinite) in the Hilbert space $\mathbb{C}^{n_1}\otimes\mathbb{C}^{n_2}\cdots \otimes \mathbb{C}^{n_m}$ is separable if and only if $\rho$ is a combination of product states, that is
\begin{equation*}
\label{msep}
\rho=\sum\limits_{k} p_k \rho_k^1\otimes \rho_k^2\otimes \cdots \otimes \rho_k^m,
\end{equation*}
where $p_k$ is positive, and $\rho_k^i$ is the state of the $i$th subsystem. If all the states $\rho,\rho_k^i$ are normalized, i.e., tr$(\rho)=1$ and tr$(\rho_k^i)=1$, then $\{p_k\}$ is a probability distribution. The state $\rho$ can be represented as a matrix with an $m$-level nested block structure \cite{Gurvits2003-2}. It is denoted by $\rho\in \mathbb{B}(n_1,\cdots,n_m)$ with
\[\rho=\left( {\begin{array}{*{20}{c}}
   {{\rho ^{11}}} &  \cdots  & {{\rho ^{1{n_1}}}}  \\
    \cdots  &  \cdots  &  \cdots   \\
   {{\rho ^{{n_1}1}}} &  \cdots  & {{\rho ^{{n_1}{n_1}}}}  \\
\end{array}} \right),\rho^{ij}\in \mathbb{B}(n_2,\cdots,n_m),\]
where $\mathbb{B}(n_m)=\mathbb{C}^{n_m\times n_m}$. We now define a nonnegative real function on  $\mathbb{B}(n_1,n_2)$:
\[
\left\|X\right\|_{n_1,n_2}=\left\|\left(||X^{ij}||_\infty\right)\right\|_\infty, \;\; \forall \;\; X\in \mathbb{B}(n_1,n_2).
\]
In fact, this function is a matrix norm. The positivity, positive scalability and triangular inequality are easily verified. We now prove the consistency. For any $X,Y\in \mathbb{B}(n_1,n_2)$,
\begin{align*}
&\left\|XY\right\|_{n_1,n_2}=\left\|\left(||\sum\nolimits_{k=1}^{n_1}X^{ik}Y^{kj}||_\infty\right)\right\|_\infty\le \left\|\left(\sum\nolimits_{k=1}^{n_1}||X^{ik}||_\infty||Y^{kj}||_\infty\right)\right\|_\infty\\
&=\left\|\left(||X^{ij}||_\infty\right)\left(||Y^{ij}||_\infty\right)\right\|_\infty\le \left\|\left(||X^{ij}||_\infty\right)\right\|_\infty \left\|\left(||Y^{ij}||_\infty\right)\right\|_\infty=||X||_{n_1,n_2}||Y||_{n_1,n_2}.
\end{align*}

Since the spectral norm is bounded above by the Frobenius norm, it is not difficult to get
\begin{equation}
\label{normin}
||X||_{n_1,n_2}=||(||X^{ij}||_\infty)||_\infty\le ||(||X^{ij}||_2)||_2=||X||_2.
\end{equation}

Recursively, we can define  the matrix norm  $\left\|\centerdot\right\|_{n_1,\cdots,n_m}$ on $ \mathbb{B}(n_1,\cdots, n_m)$:
\[
\left\|X\right\|_{n_1,\cdots,n_m}=\left\|\left(||X^{ij}||_{n_2,\cdots,n_m}\right)\right\|_\infty, \;\; \forall \;\;X\in \mathbb{B}(n_1,\cdots,n_m).
\]
In the following, this norm will be said to be the $(n_1,\cdots, n_m)$-nested norm.

We now derive some properties of the new defined norm, which will be used in the next section.
\\
\\
\textbf{Proposition 2.1.} \begin{description}
                      \item[(i)]For any $X\in \mathbb{B}(n_1,n_2)$,
\[
\left\|X\right\|_\infty\le \left\|X\right\|_{n_1,n_2}\le n_1\left\|X\right\|_\infty.
\]
                      \item[(ii)] For any $X\in \mathbb{C}^{n\times n}$,
                      \[
                      \left\|X\right\|_\infty\ge \max\left\{ \left\|H(X)\right\|_\infty,\; \left\|S(X)\right\|_\infty\right\}.
                      \]
                      \item[(iii)] For any $X\in \mathbb{B}(n_1,n_2)$,
                      \[
                      \left\|X\right\|_{n_1,n_2}\ge \max\left\{ \left\|H(X)\right\|_{n_1,n_2},\; \left\|S(X)\right\|_{n_1,n_2}\right\}.
                      \]
                    \end{description}
\textbf{Proof.} \emph{Proof of }(i). The first inequality is due to the proof of \cite[Theorem 1]{Gurvits2002}.
From \cite[Corollary 3.8]{Stewart1990}, we have
\[\left\|X\right\|_\infty\ge \max\limits_{1\le i,j\le n_1} \left\|X^{ij}\right\|_\infty,
\]
which implies, by setting $e=(1,\cdots,1)^\dag\in \mathbb{C}^{n_1}$,
\[
\left\|X\right\|_{n_1,n_2}=\left\|\left(||X^{ij}||_\infty\right)\right\|_\infty\le \left\|X\right\|_\infty ||ee^\dag||_\infty=n_1\left\|X\right\|_\infty.
\]

\emph{Proof of} (ii). From \cite[Theorem 1.2]{Goldberg1982}, we can get
\begin{align*}
 \left\|X\right\|_\infty&\ge \max\limits_{||x||_\infty=1,x\in \mathbb{C}^{n}}|x^\dag Xx|=\max\limits_{||x||_\infty=1,x\in \mathbb{C}^{n}}|x^\dag H(X)x+x^\dag S(X)x|.
\end{align*}
Since $x^\dag H(X)x$ is real, and $x^\dag S(X)x$ is pure imaginary or zero, we further have
\begin{align*}
\left\|X\right\|_\infty&\ge \max\left\{\max\limits_{||x||_\infty=1,x\in \mathbb{C}^{n}}|x^\dag H(X)x|,\max\limits_{||x||_\infty=1,x\in \mathbb{C}^{n}}|x^\dag S(X)x|\right\}\\
&=\max\left\{ \left\|H(X)\right\|_\infty,\; \|S(X)\|_\infty\right\}.
\end{align*}

\emph{Proof of }(iii). On one hand, by (ii),
\begin{align*}
 \left\|X\right\|_{n_1,n_2}&= \left\|\left(||X^{ij}||_{\infty}\right)\right\|_{\infty}\ge \left\|\left(\frac{1}{2}(||X^{ij}||_\infty+||(X^{ji})^\dag||_\infty)\right)\right\|_{\infty}\\
 &\ge \left\|\left(\frac{1}{2}||X^{ij}+(X^{ji})^\dag||_\infty\right)\right\|_{\infty}=\left\|H(X)\right\|_{n_1,n_2}.
\end{align*}
On the other hand, applying the result given above to $\text{i}X=\text{i}H(X)+\text{i}S(X)$ with $\text{i}=\sqrt{-1}$ yields
\begin{align*}
 \left\|X\right\|_{n_1,n_2}&=\left\|\text{i}X\right\|_{n_1,n_2}\ge \left\|H(\text{i}X)\right\|_{n_1,n_2}=\left\|\text{i}S(X)\right\|_{n_1,n_2}=\left\|S(X)\right\|_{n_1,n_2}.
\end{align*}
The proof is completed. $\hfill \Box$

\section{The separable ball for the bipartite quantum system and its applications}

It was shown in \cite{Gurvits2002} that the unnormalized state $I_{n_1n_2}+\eta$ in $\mathbb{C}^{n_1}\otimes \mathbb{C}^{n_2}$ is separable for all $\eta$ satisfying $||\eta||_2\le 1$. In this section, the condition $||\eta||_2\le 1$ is improved to be $||\eta||_{n_1,n_2}\le 1$. The following lemma is due to \cite{Woronowicz-Horodecki}; see also \cite{Gurvits2002}.
\\
\\
\textbf{Lemma 3.1.} The unnormalized state $\rho$ in $\mathbb{C}^{n_1}\otimes \mathbb{C}^{n_2}$ is separable if and only if, for any positive stochastic linear operator (i.e., preserving positive semidefiniteness and identity matrices) $\Lambda_2:\mathbb{C}^{n_2\times n_2}\rightarrow \mathbb{C}^{n_2'\times n_2'}$, $(I_{n_1}\otimes \Lambda_2)\rho$ is positive semidefinite.
\\
\\
\textbf{Theorem 3.1.} The unnormalized state $\rho$ in $\mathbb{C}^{n_1}\otimes \mathbb{C}^{n_2}$ is separable if $||\rho-I_{n_1n_2}||_{n_1,n_2}\le 1$.
\\
\\
\textbf{Proof.} Set $\delta=\rho-I_{n_1n_2}$. For any positive stochastic linear operator $\Lambda_2:\mathbb{C}^{n_2\times n_2}\rightarrow \mathbb{C}^{n_2\times n_2}$, from the proof of \cite[Theorem 1 and Proposition 2]{Gurvits2002} we have
\[
\left\|(I_{n_1}\otimes \Lambda_2)\delta\right\|_\infty\le \left\|\left(\left\|\Lambda_2(\delta^{ij})\right\|_\infty\right)\right\|_\infty\le \left\|\left(\left\|\delta^{ij}\right\|_\infty\right)\right\|_\infty=\left\|\delta\right\|_{n_1,n_2}\le 1,\]
which implies that $(I_{n_1}\otimes \Lambda_2)\rho=I_{n_1n_2}+(I_{n_1}\otimes \Lambda_2)\delta$ is positive semidefinite. Thus, by Lemma 3.1 $\rho$ is separable. $\hfill \Box$\\

Since $\rho$ can have negative eigenvalues if $||\rho-I_{n_1n_2}||_{n_1,n_2}> 1$, we get that the ball given by Theorem 3.1 is the largest one with the new norm.  If the normalized state $\rho$ in $\mathbb{C}^{n_1}\otimes \mathbb{C}^{n_2}$, i.e., tr($\rho$)=1, satisfies
$$||\rho-\frac{1}{n_1n_2}I_{n_1n_2}||_{n_1,n_2}\le \frac{1}{n_1n_2},$$
then by Theorem 3.1 $\rho$ is separable.

From (\ref{normin}), it follows that the ball given by Theorem 3.1 is larger than the one showed by \cite[Theorem 1]{Gurvits2002}. For example, consider the $d$-dimensional Werner states \cite{Werner1989-2}

\[
\rho_W=\frac{1}{d^3-d} \left((d-b)I_{d^2}+(db-1)\eta\right),
\]
where $-1\le b\le 1$, the ``flip" or ``swap" operator $\eta$ can be represented as $\eta=\sum\nolimits_{i,j=0}^{d-1}|ij\rangle\langle ji|$, and $\rho_W$ is separable if and only if $0\le b\le 1$.
Since
\[
||d^2\rho_W-I_{d^2}||_{d,d}=|1-db|\text{ and }||d^2\rho_W-I_{d^2}||_2=\frac{d|1-db|}{\sqrt{d^2-1}},
\]
due to Theorem 3.1 and \cite[Theorem 1]{Gurvits2002}, we get the separable conditions, respectively,
\begin{equation}
\label{firstin}
0\le b\le \frac{2}{d}\text{ and }\frac{d-\sqrt{d^2-1}}{d^2}\le b\le \frac{d+\sqrt{d^2-1}}{d^2}.
\end{equation}
Clearly, the former is better than the later. Moreover, Theorem 1 becomes sufficient and necessary for the separability of the $2$-dimensional Werner state.

By the parameter scaling technique and Theorem 3.1, the following slightly better separable condition for Werner state can be obtained:
\[
0\le b\le \frac{2d-1}{d^2-d+1}.
\]
In fact, we only consider the case $\frac{2}{d}\le b\le \frac{2d-1}{d^2-d+1}$ from (\ref{firstin}). Taking $a=\frac{2d^2-d-2db+d^2b}{2(d^2-1)}$ yields
\[
(d^2-db-d^2a+a)+d^2b-d= -(d^2-db-d^2a+a)\ge 0,
\]
and, furthermore,
\begin{align*}
\left\|\frac{d^2}{a}\rho_W-I_{n_1n_2}\right\|_{n_1,n_2}&=\frac{1}{a(d^2-1)}\left((d^2-db-d^2a+a+d^2b-d)+(d-1)(d^2b-d)\right)\\
&=\frac{1}{a(d^2-1)}(d^3b-db-d^2a+a)\le 1.
\end{align*}
Therefore, $\rho_W$ is separable from Theorem 3.1. Similar analysis can be used for the isotropic states \cite{isotropic}.

 The following well-known $3\times 3$ bound entangled state was due to Horodecki \cite{Horodecki1997}:
\[\rho=\frac{1}{8a+1}\left( {\begin{array}{*{20}{c}}
   {a} & {0} & {0} & {0} & {a} & {0} & {0} & {0} & {a}  \\
   {0} & {a} & {0} & {0} & {0} & {0} & {0} & {0} & {0}  \\
   {0} & {0} & {a} & {0} & {0} & {0} & {0} & {0} & {0}  \\
   {0} & {0} & {0} & {a} & {0} & {0} & {0} & {0} & {0}  \\
   {a} & {0} & {0} & {0} & {a} & {0} & {0} & {0} & {a}  \\
   {0} & {0} & {0} & {0} & {0} & {a} & {0} & {0} & {0}  \\
   {0} & {0} & {0} & {0} & {0} & {0} & {\frac{1+a}{2}} & {0} & {\frac{\sqrt{1-a^2}}{2}}  \\
   {0} & {0} & {0} & {0} & {0} & {0} & {0} & {a} & {0}  \\
   {a} & {0} & {0} & {0} & {a} & {0} & {\frac{\sqrt{1-a^2}}{2}} & {0} & {\frac{1+a}{2}}  \\
\end{array}} \right),\]
where $0<a<1$. We consider the mixture of this state with the maximally mixed state:
$$
\rho_{mix}=p\rho +\frac{1-p}{9}I_9,
$$
 where $0\le p\le 1$. The separable conditions of $\rho_{mix}$ from \cite[Theorem 1]{Gurvits2002} and Theorem 3.1 are displayed in Table 1 for different values of $a$. It is easy to be found that Theorem 3.1 is more efficient than \cite[Theorem 1]{Gurvits2002}.

 \begin{table}[htbp]
\centering \begin{tabular} {c|c|c}\hline
\raisebox{-1.50ex}[0cm][0cm]{$a=0.25$}& \cite[Theorem 1]{Gurvits2002}& $0\le p\le 0.3233$ \\ \cline{2-3}
 &
Theorem 3.1& $0\le p\le 0.3430$
 \\ \hline \raisebox{-1.50ex}[0cm][0cm]{$a=0.50$}& \cite[Theorem 1]{Gurvits2002}& $0\le p\le 0.3955$
  \\ \cline{2-3}
 &
Theorem 3.1& $0\le p\le 0.4275$ \\ \hline \raisebox{-1.50ex}[0cm][0cm]{$a=0.75$}& \cite[Theorem 1]{Gurvits2002}&$0\le p\le 0.4089$ \\ \cline{2-3}
&
Theorem 3.1& $0\le p\le 0.4635$ \\ \hline
\end{tabular} \caption{\emph{Separable conditions of $\rho_{mix}$ from \cite[Theorem 1]{Gurvits2002} and Theorem 3.1}}
\end{table}

If the unnormalized state $\rho$ in $\mathbb{C}^{n_1}\otimes \mathbb{C}^{n_2}$ satisfies $||\rho-I_{n_1n_2}||_\infty \le \frac{1}{n_1}$, then, from Proposition 2.1(i) and Theorem 3.1, $\rho$ is separable. This is a welcome separable condition, since determining experimentally the spectrum of the unknown density matrix can be easier than constructing the full density matrix \cite{Tanaka2014}. Nevertheless, by \cite[Corollary 1]{Gurvits2002}, we can get the same separable condition.\\
\\
\emph{Applications for pseudopure states.}
The class of pseudopure states plays a crucial role in nuclear magnetic resonance quantum information processing; see, e.g., \cite{Gurvits2003-2,pseudopure}. In the following, we consider the separability of the normalized pseudopure state in $\mathbb{C}^{n_1}\otimes\mathbb{C}^{n_2}$:
\begin{equation}
\label{pps}
\rho_{\pi,\varepsilon}=\varepsilon \pi+\frac{1-\varepsilon}{n_1n_2}I_{n_1n_2},
\end{equation}
where $\varepsilon \ge 0$, and $\pi$ is a normalized pure state.
\\
\\
\textbf{Corollary 3.1.} Let $\rho_{\pi,\varepsilon}$ be defined as in (\ref{pps})
with
\[
\pi=\left( {\begin{array}{*{20}{c}}
   {{X_1}}  \\
   {{\vdots}}\\
   {{X_{n_1}}}  \\
\end{array}} \right)\left( {\begin{array}{*{20}{c}}
   {X_1^\dag} &{\cdots}& {X_{n_1}^{\dag}}  \\
\end{array}} \right), X_i\in \mathbb{C}^{n_2},i=1,\cdots,n_1.
\]
If
\begin{equation}
\label{vps}
\varepsilon\le \min\limits_{1\le i\le n_1}\left\{\frac{1}{n_1n_2(1-||X_i||^2_\infty)+1}\right\},
\end{equation}
then $\rho$ is separable.
\\
\\
\textbf{Proof.} The state $\rho_{\pi,\varepsilon}$ can be written as, for any $a>0$,
 \begin{equation}
 \label{spp} \rho_{\pi,\varepsilon}=\varepsilon a \left(I_{n_1n_2}+\frac{1}{a}\tau\right),
 \end{equation}
 where
 \begin{equation}
 \label{sigmab}\tau=\pi-(a-b) I_{n_1n_2},b=\frac{1-\varepsilon}{n_1n_2\varepsilon}.
 \end{equation}
Setting $a=\frac{1}{2}(1+b)$, we get by (\ref{vps}), for $1\le i\le n_1$,
\[
||X_i||_\infty^2\ge \frac{\varepsilon n_1n_2+\varepsilon-1}{\varepsilon n_1n_2}=2(a-b),
\]
and then
\[
||X_iX_i^\dag-(a-b)I_{n_2}||_\infty=||X_i||_\infty^2-(a-b).
\]
Thus, it follows that
\begin{align*}
 \left\|\tau\right\|_{n_1,n_2}&={\left\| {\begin{array}{*{20}{c}}
   {||X_1X_1^\dag-(a-b)I_{n_2}||_\infty } &\cdots& {||{X_1}{X_{n_1}^\dag}|{|_\infty }}  \\
   \cdots&\cdots & \cdots&\\
   {||{X_{n_1}}{X_1^\dag}|{|_\infty }} &\cdots& {||{X_{n_1}X_{n_1}^\dag}-(a-b)I_{n_2}||_\infty }  \\
\end{array}} \right\|_\infty }\\
&={\left\| {\begin{array}{*{20}{c}}
   {||X_1||^2_\infty-(a-b) } &\cdots& {||{X_1}||_\infty {||X_{n_1}||_\infty} }  \\
  \cdots &\cdots & \cdots&\\
   {||{X_1}||_\infty {||X_{n_1}||_\infty}} &\cdots& {||X_{n_1}||^2_\infty-(a-b) }  \\
\end{array}} \right\|_\infty }\\
&=\left\|\left( {\begin{array}{*{20}{c}}
   {{||X_1||_\infty}}  \\
   {{\vdots}}\\
   {{||X_{n_1}||_\infty}}  \\
\end{array}} \right)\left( {\begin{array}{*{20}{c}}
   {||X_1||_\infty} &{\cdots}& {||X_{n_1}||_\infty}  \\
\end{array}} \right)-(a-b)I_{n_1}\right\|_\infty\\
&=\max\{|1-a+b|,|a-b|\}=a.
 \end{align*}
Hence, by (\ref{spp}) and Theorem 3.1, $\rho_{\pi,\varepsilon}$ is separable. $\hfill \Box$\\

It was shown in \cite[Section IV]{Gurvits2003-2} that $\rho_{\pi,\varepsilon}$ is separable for $\varepsilon\le \frac{1}{\sqrt{n_1n_2(n_1n_2-1)}}$. This condition is worse than (\ref{vps}) under the case \[
\min\limits_{1\le i\le n_1}\{||X_i||_\infty^2\}\ge\frac{1}{n_1n_2}\left(n_1n_2+1-\sqrt{n_1n_2(n_1n_2-1)}\right).
\]
For $n_1=n_2=N$, the separable condition (\ref{vps}) is better than the one $\varepsilon\le \frac{1}{N^2-1}$ given by \cite[Corollary 5]{Gurvits2002} if $\min\nolimits_{1\le i\le n_1}\{||X_i||_\infty^2\}\ge \frac{2}{N^2}$.

It follows from Corollary 3.1 that if $\min\nolimits_{1\le i\le n_1}\{||X_i||^2_\infty\}$ sufficiently approximates to zero, then the upper bound of $\varepsilon$ is sufficiently close to $\frac{1}{n_1n_2+1}$. This upper bound is not ideal. For $n_1=2$, we can get a better separable condition.
\\
\\
\textbf{Corollary 3.2.} Let $\rho_{\pi,\varepsilon}$ be defined as in (\ref{pps}) with $n_1=2$ and
\[
\pi=\left( {\begin{array}{*{20}{c}}
   {{X_1}}  \\
   {{X_2}}  \\
\end{array}} \right)\left( {\begin{array}{*{20}{c}}
   {X_1^\dag} & {X_2^{\dag}}  \\
\end{array}} \right), X_1,X_2\in \mathbb{C}^{n_2}.
\]
If $\varepsilon\le \frac{\sqrt{3}}{2n_2+\sqrt{3}}$, then $\rho$ is separable.
\\
\\
 \textbf{Proof.} From $||X_1||_\infty^2+||X_2||_\infty^2=1$, without loss of generality, we assume $||X_2||_\infty^2\le \frac{1}{2}$. Therefore, \begin{equation}
 \label{cor3.21}\frac{1}{2}\le ||X_1||_\infty^2\le 1.
 \end{equation}
  Let $a,b $ and $\tau$ be defined as in  (\ref{spp}) and (\ref{sigmab}) with $n_1=2$. Then setting $a=b+\frac{1}{4}$ results in, by $\varepsilon\le \frac{\sqrt{3}}{2n_2+\sqrt{3}}$,
 \begin{equation}
\label{a} a=\frac{1-\varepsilon}{2n_2\varepsilon}+\frac{1}{4}\ge \frac{1}{4}+\frac{\sqrt{3}}{3}.
\end{equation}
Defining the function
  \begin{align*}
 f(y)&=\frac{1}{2}\left(y+\sqrt{(y-\frac{1}{2})^2+4y(1-y)}\right),\quad y\in \mathbb{C},
  \end{align*}
 we derive, from (\ref{cor3.21}),
 \begin{align}
 \left\|\tau\right\|_{2,n_2}&={\left\| {\begin{array}{*{20}{c}}
  ||X_1X_1^{\dag}-\frac{1}{4}I_{n_2}||_\infty & ||X_1X_2^{\dag}||_\infty  \\
  ||X_2X_1^{\dag}||_\infty  & ||X_2X_2^{\dag}-\frac{1}{4}I_{n_2}||_\infty \\
\end{array}} \right\|_\infty }\nonumber\\
&={\left\| {\begin{array}{*{20}{c}}
   {||{X_1}||_\infty ^2 - \frac{1}{{4}}} & {||{X_1}|{|_\infty }||{X_2}|{|_\infty }}  \\
   {||{X_1}|{|_\infty }||{X_2}|{|_\infty }} & {\frac{1}{{4}}}  \\
\end{array}} \right\|_\infty }\nonumber\\
\label{cor322} &=\frac{1}{2}\left(||X_1||_\infty^2+\sqrt{\left(||X_1||_\infty^2-\frac{1}{2}\right)^2+4||X_1||_\infty^2||X_2||_\infty^2}\right)
 =f(||X_1||^2_\infty).
  \end{align}
 It follows from (\ref{cor3.21}), (\ref{a}) and (\ref{cor322}) that
  \[
  \left\|\tau\right\|_{2,n_2}=f\left(||X_1||^2_\infty\right)\le \max\limits_{\frac{1}{2}\le y\le 1}\{f(y)\}=f\left(\frac{3+\sqrt{3}}{6}\right)=\frac{1}{4}+\frac{\sqrt{3}}{3}\le a.
  \]
 Thus, due to (\ref{spp}) and Theorem 3.1 $\rho_{\pi,\varepsilon}$ is separable. $\hfill \Box$\\

It is easy to be verified that the separable condition $\varepsilon\le \frac{\sqrt{3}}{2n_2+\sqrt{3}}$ is always better than the condition $\varepsilon\le \frac{1}{\sqrt{2n_2(2n_2-1)}}$ given by \cite[Section IV]{Gurvits2003-2}.
\\
\\
\emph{Applications for states with strong positive partial transposes.}
In \cite{Chruscinski2008}, Chru\'{s}ci\'{n}ski, Jurkowski and Kossakowski proposed a class of unnormalized bipartite states with strong positive partial transposes by considering the block structure of the density matrix. They are said to be SPPT states, and cover many previously known separable states. It was shown in \cite{Ha} that not all the SPPT states are separable. We now recall the definition of SPPT states in $\mathbb{C}^{2}\otimes \mathbb{C}^{n_2}$.

Consider the following class of unnormalized states in $\mathbb{C}^2\otimes \mathbb{C}^{n_2}:$
\begin{equation}
\label{rho}
\rho =X^{\dag}X \text{ with }X=\left( {{\begin{array}{*{20}c}
X_1& {SX_1}  \\
 {0} &X_2 \\
\end{array} }} \right),
\end{equation}
where $X_1,X_2$ and $S$ are arbitrary complex $n_2\times n_2$ matrices. The unnormalized state $\rho$ is said to be SPPT if and only if $X_1^\dag S^\dag S X_1=X_1^\dag S S^\dag X_1$. It was found in \cite{Bylicka2013} that, if $S$ is a normal matrix, then the SPPT state $\rho$ is separable.

Set  $\sigma_{\min}$ and $\sigma_{\max}$ are the smallest and largest singular values of $S$, respectively. In the following, we will give some separable conditions for $\rho$, but $\rho$ is not necessarily SPPT.
\\
\\
\textbf{Corollary 3.3.} Let $\rho$ be an unnormalized state defined as in (\ref{rho}). Then if $X_2^\dag X_2-X_1^\dag X_1$ is positive semidefinite, and $\sigma_{\max}^2-\sigma_{\min}^2\le 1$, then $\rho$ is separable.
\\
\\
\textbf{Proof.} The state $\rho$ can be decomposed into
\begin{align*}
\rho&=\left( {{\begin{array}{*{20}c}
X_1^{\dag}& {0}  \\
 {0} &X_1^{\dag} \\
\end{array} }} \right)\left( {{\begin{array}{*{20}c}
I_{n_2}& {S}  \\
 {S^{\dag}} &I_{n_2}+S^{\dag} S \\
\end{array} }} \right)\left( {{\begin{array}{*{20}c}
X_1& {0}  \\
 {0} &X_1 \\
\end{array} }} \right)+\left( {{\begin{array}{*{20}c}
0& 0  \\
 {0} & X_2^{\dag} X_2-X_1^{\dag} X_1\\
\end{array} }} \right)\\
&:=P^{\dag} WP+V.
\end{align*}
Clearly, $V$ is separable, since it can be represented as
\[
\left( {{\begin{array}{*{20}c}
0& {0}  \\
 {0} &1\\
\end{array} }} \right)\otimes (X_2^\dag X_2-X_1^\dag X_1).\]
 We now prove that $P^\dag W P$ is separable. In fact, it is easy to deduce
\[
W-(1+\sigma_{\max}^2)I_{2n_2}=(1+\sigma_{\max}^2)\left(\frac{1}{1+\sigma_{\max}^2}W-I_{2n_2}\right),
\]
and, by $||S^{\dag}S-\sigma_{\max}^2I_{n_2}||_\infty=\sigma_{\max}^2-\sigma_{\min}^2$ and $\sigma_{\max}^2-\sigma_{\min}^2\le 1$,
\begin{align*}
\left\| \frac{1}{1+\sigma_{\max}^2}W-I_{2n_2}\right\|_{2,n_2}&=\frac{1}{1+\sigma_{\max}^2}\left\| {{\begin{array}{*{20}c}
\sigma_{\max}^2& {\sigma_{\max}}  \\
 {\sigma_{\max}} &\sigma_{\max}^2-\sigma_{\min}^2 \\
\end{array} }}\right\|_\infty\\
&=\frac{1}{2(1+\sigma_{\max}^2)}\left(2\sigma_{\max}^2-\sigma_{\min}^2+\sqrt{\sigma_{\min}^4+ 4\sigma_{\max}^2}\right)\le 1.
\end{align*}
Hence, from Theorem 3.1, $\frac{1}{1+\sigma_{\max}^2}W$ is separable, and, furthermore, $P^\dag WP$ is separable. Thus, $\rho$ is separable. $\hfill \Box$\\

 If $X_1$ is nonsingular, then we denote by $\eta_{\min}$ and $\eta_{\max}$ the smallest and largest eigenvalues of $R:=S^\dag S+X_1^{-\dag}X_2^\dag X_2 X_1^{-1}$, respectively, and then obtain the separable conditions as follows.\\
 \\
\textbf{Corollary 3.4.} Let $\rho$ be an unnormalized state defined as in (\ref{rho}), and let $X_1$ be nonsingular.
Then if one of the following conditions
\begin{description}
  \item[(i)] $\eta_{\min}\ge \sigma^2_{\max}, \eta_{\max}\ge 1;$
  \item[(ii)] $\eta_{\max}\le 1,  \eta_{\min}(2\eta_{\max}-1)\ge \sigma^2_{\max}$;
  \item[(iii)] $1\ge \eta_{\min}\ge \frac{1}{4}(1+\eta_{\max}),
  (2\eta_{\min}-1)(2\eta_{\min}-\eta_{\max})\ge \sigma_{\max}^2$;
\end{description}
holds, then $\rho$ is separable.
\\
\\
 \textbf{Proof.} Since $X_1$ is nonsingular, $\rho$ can be written as
 \[
 \rho=\left( {{\begin{array}{*{20}c}
X_1^\dag& {0}  \\
 {0} &X_1^\dag \\
\end{array} }} \right)\left( {{\begin{array}{*{20}c}
I_{n_2}& {S}  \\
 {S^\dag} &S^\dag S+X_1^{-\dag}X_2^\dag X_2X_1^{-1} \\
\end{array} }} \right)\left( {{\begin{array}{*{20}c}
X_1& {0}  \\
 {0} &X_1 \\
\end{array} }} \right):=Q^\dag Y Q.
\]
By $Y-aI_{2n_2}=a(\frac{1}{a}Y-I_{2n_2})$ with $a>0$ and Theorem 3.1, we only need to prove that there exists $a>0$ such that $||\frac{1}{a}Y-I_{2n_2}||_{2,n_2}\le 1$, i.e.,
\begin{equation}
\label{trans}
||Y-a I_{2n_2}||_{2,n_2}=\left\| {{\begin{array}{*{20}c}
|1-a|& {\sigma_{\max}}  \\
 {\sigma_{\max}} & ||R-aI_{n_2}||_\infty  \\
\end{array} }} \right\|_\infty \le a.
\end{equation}

\emph{Proof of }(i). In this case, taking $a=\eta_{\max}$, we get $a\ge 1$, and then
\begin{align*}
||Y-\eta_{\max} I_{2n_2}||_{2,n_2}&=\left\| {{\begin{array}{*{20}c}
\eta_{\max}-1& {\sigma_{\max}}  \\
 {\sigma_{\max}} & \eta_{\max}-\eta_{\min}  \\
\end{array} }} \right\|_\infty\\
&=\frac{1}{2}\left(2\eta_{\max}-\eta_{\min}-1+\sqrt{(\eta_{\min}-1)^2+4\sigma_{\max}^2}\right),
\end{align*}
which, together with $\eta_{\min}\ge \sigma_{\max}^2$, implies $||Y-\eta_{\max} I_{2n_2}||_{2,n_2}\le \eta_{\max}=a$.

\emph{Proof of }(ii). In this case, setting $a=\eta_{\max}$ yields $a\le 1$, and then
\begin{align*}
||Y-\eta_{\max} I_{2n_2}||_{2,n_2}&=\left\| {{\begin{array}{*{20}c}
1-\eta_{\max}& {\sigma_{\max}}  \\
 {\sigma_{\max}} & \eta_{\max}-\eta_{\min}  \\
\end{array} }} \right\|_\infty\\
&=\frac{1}{2}\left(1-\eta_{\min}+\sqrt{(2\eta_{\max}-\eta_{\min}-1)^2+4\sigma_{\max}^2}\right).
\end{align*}
Thus, by $\eta_{\min}(2\eta_{\max}-1)\ge \sigma^2_{\max}$, we can get $||Y-\eta_{\max} I_{2n_2}||_{2,n_2}\le \eta_{\max}=a$.

\emph{Proof of }(iii). In this case, taking $a=\eta_{\min}$, we get $a\le 1$, and then
\begin{align*}
||Y-\eta_{\min} I_{2n_2}||_{2,n_2}&=\left\| {{\begin{array}{*{20}c}
1-\eta_{\min}& {\sigma_{\max}}  \\
 {\sigma_{\max}} & \eta_{\max}-\eta_{\min}  \\
\end{array} }} \right\|_\infty\\
&=\frac{1}{2}\left(1+\eta_{\max}-2\eta_{\min}+\sqrt{(\eta_{\max}-1)^2+4\sigma_{\max}^2}\right),
\end{align*}
which, from $\eta_{\min}\ge \frac{1}{4}(1+\eta_{\max})$ and $
  (2\eta_{\min}-1)(2\eta_{\min}-\eta_{\max})\ge \sigma_{\max}^2$, implies $||Y-\eta_{\min} I_{2n_2}||_{2,n_2}\le \eta_{\min}=a$. $\hfill \Box$\\

   Obviously, Corollaries 3.3 and 3.4 are not limited to be for SPPT states. If the $n_1n_2$ times of a normalized state $\rho$ in $\mathbb{C}^{n_1}\otimes \mathbb{C}^{n_2}$ satisfies the conditions of Corollary 3.3 (respectively, Corollary 3.4($l$), $l=$(i),(ii),(iii)), we denote it by $\rho\in\mathcal{A}$ (respectively, $\rho\in \mathcal{B}_l,l=$(i),(ii),(iii)). The sets $\mathcal{A}$ and $\mathcal{B}_l,l=$(i),(ii),(iii) must contain some states that are not SPPT. We now show that, similar to the SPPT states, these sets of  states also cover some well-known separable states.

    The maximally mixed state clearly belongs to $\mathcal{A}$ and $\mathcal{B}_l,l=$(i),(ii),(iii) with $X_1=X_2=I_{n_2}$ and $S=0$. Thus, both $\mathcal{A}$ and $\mathcal{B}_l,l=$(i),(ii),(iii) include the SPPT Werner states \cite{Werner1989-2} and isotropic states  \cite{isotropic} in $\mathbb{C}^2\otimes \mathbb{C}^{2}$, since these states are maximally mixed (\cite{Chruscinski2008}). In \cite{Horodecki1997}, the author constructed a PPT state in $\mathbb{C}^2\otimes \mathbb{C}^4$ with a parameter $b\in [0,1]$.  This PPT state is SPPT if and only if $b=0$ (\cite{Chruscinski2008}). From Corollaries 3.3 and 3.4, this PPT state with $b=0$ belongs to $\mathcal{A}$ with $X_1=S=0$, while this PPT state with $b=1$ is contained in $\mathcal{B}_{l}$, $l$=(i),(ii),(iii) with $X_1=I_{n_2}$.

Consider the normalized circulant PPT sates in $\mathbb{C}^2\otimes \mathbb{C}^2$ \cite{Chruscinski2007}
\[\rho=\left( {\begin{array}{*{20}{c}}
   {{a_{11}}} & 0 & 0 & {{a_{12}}}  \\
   0 & {{b_{11}}} & {{b_{12}}} & 0  \\
   0 & {{b_{21}}} & {{b_{22}}} & 0  \\
   {{a_{21}}} & 0 & 0 & {{a_{22}}}  \\
\end{array}} \right).
\]
By \cite{Chruscinski2008}, $\rho$ is a SPPT state if and only if $|a_{12}|=|b_{12}|$. In fact, this SPPT state can be contained in $\mathcal{A}$ and $\mathcal{B}_{l}$ under some conditions. For example, if either $b_{22}\ge a_{11}>0, b_{11}>0$ or $a_{22}\ge b_{11}>0, a_{11}>0$ holds, then $\rho\in \mathcal{B}_{\text{(i)}}$.
\\
\\
\emph{Applications for $m$-partite quantum system.}
Theorem 3.1 will be used to provide the separable ball for multipartite quantum system. The following lemmas are needed. \\
\\
\textbf{Lemma 3.2}\cite{Gurvits2003-2}. Let $C(n_2)\subset \mathbb{H}\mathbb{C}^{n_2\times n_2}$ be a cone. Then $X$ is $\mathbb{P}\mathbb{C}^{n_1\times n_1}\otimes C(n_2)$ separable if and only if $(I_{n_1}\otimes \Lambda)X$ is positive semidefinite for all $C(n_2)$-positive linear operators $\Lambda: \mathbb{C}^{n_2\times n_2}\rightarrow \mathbb{C}^{n_1\times n_1}.$

Similar to the definition of the cone given in \cite[Definition 6]{Gurvits2003-2}, we denote by $\mathbb{Z}(n_1,\cdots,n_m;a)$ the cone generated by Hermitian matrices of the form \[
\left\{I_{n_1\cdots n_m}+\vartriangle: ||\vartriangle||_{n_1,\cdots,n_m}\le a, \vartriangle\in \mathbb{B}(n_1,\cdots,n_m) \right\}.
\]
\textbf{Lemma 3.3.} Let $\Lambda: \mathbb{C}^{n_1n_2\times n_1n_2}\rightarrow \mathbb{C}^{n_3\times n_3}$ be a $\mathbb{Z}(n_1,n_2;a)$-positive linear operator. Then
\begin{description}
  \item[(i)] for any Hermitian $X\in \mathbb{B}(n_1,n_2),$
  \[
  \left\|\Lambda(X)\right\|_\infty\le \frac{1}{a}  \left\|X|\right\|_{n_1,n_2}.
  \]
  \item[(ii)] for any  $Y\in \mathbb{B}(n_1,n_2)$,
  \[
  \left\|\Lambda(Y)\right\|_\infty\le \frac{2}{a}  \left\|Y|\right\|_{n_1,n_2}.
  \]
\end{description}
\textbf{Proof.} \emph{Proof of }(i). By
\[
\left\|\frac{aX}{||X||_{n_1,n_2}}\right\|_{n_1,n_2}=a,
\]
 we get $\Lambda\left(I_{n_1n_2}+\frac{aX}{||X||_{n_1,n_2}}\right)$ is positive semidefinite, and then \[
\left\|\Lambda\left(\frac{aX}{||X||_{n_1,n_2}}\right)\right\|_\infty\le 1,\]
which implies that the conclusion holds.

\emph{Proof of }(ii). From (i) and Proposition 2.1, it deduces
\[
\left\|\Lambda(Y)\right\|_\infty\le \left\|\Lambda(H(Y))\right\|_\infty+\left\|\Lambda(S(Y))\right\|_\infty\le \frac{1}{a}\left(\left\|H(Y)\right\|_{n_1,n_2}+\left\|S(Y)\right\|_{n_1,n_2}\right)\le \frac{2}{a}\left\|Y\right\|_{n_1,n_2}.
\]
$\hfill\Box$

The proof of the following lemma is similar to that of \cite[Theorem 1]{Gurvits2003-2}.\\
\\
\textbf{Lemma 3.4.} Let $\rho$ be an unnormalized state in $\mathbb{C}^{n_1}\otimes \mathbb{C}^{n_2}\otimes\mathbb{C}^{n_3}$. If $\left\|I-\rho\right\|_{n_1,n_2,n_3}\le \frac{a}{2}$, then $\rho$ is $\mathbb{P}\mathbb{C}^{n_1\times n_1}\otimes \mathbb{Z}(n_2,n_3;a)$ separable.
\\
\\
\textbf{Proof.} Let $\tau=I_{n_1n_2n_3}-\rho$, and $\Lambda$ be any $\mathbb{Z}(n_2,n_3;a)$-positive linear operator. Then, from Lemma 3.3 we achieve
\[
||(I_{n_1}\otimes \Lambda)\tau||_\infty\le ||(||\Lambda(\tau^{ij})||_\infty)||_\infty\le \frac{2}{a}||(||\tau^{ij}||_{n_1,n_2})||_\infty=\frac{2}{a}||\tau||_{n_1,n_2,n_3}\le 1,
\]
which implies that $(I_{n_1}\otimes \Lambda)\rho=I_{n_1n_2n_3}-(I_{n_1}\otimes \Lambda)\tau$ is positive semidefinite. Due to Lemma 3.2, $\rho$ is $\mathbb{P}\mathbb{C}^{n_1\times n_1}\otimes \mathbb{Z}(n_2,n_3;a)$ separable. $\hfill \Box$\\

We now give the separable ball for the $m$-partite quantum system. Its proof is similar to that of \cite[Corollary 1]{Gurvits2003-2}.
\\
\\
\textbf{Corollary 3.5.} If the $m$-partite unnormalized state $\rho$ in $\mathbb{C}^{n_1}\otimes \cdots \otimes\mathbb{C}^{n_m}$ satisfies $||I_{n_1\cdots n_m}-\rho||_{n_1,\cdots,n_m}\le \frac{1}{2^{m-2}}$, then $\rho$ is separable.
\\
\\
\textbf{Proof.} We prove the conclusion by mathematical induction on the number of subsystems. By Theorem 3.1, all states in $\mathbb{Z}(n_1,n_2;1)$ are separable. We now suppose that all states in $\mathbb{Z}(n_1,n_2,\cdots, n_{m-1};\frac{1}{2^{m-3}})$ are separable. Then by the condition $||I_{n_1\cdots n_m}-\rho||_{n_1,\cdots,n_m}\le \frac{1}{2^{m-2}}$ and Lemma 3.4, $\rho$ is $\mathbb{P}\mathbb{C}^{n_m}\otimes \mathbb{Z}(n_1,n_2,\cdots, n_{m-1};\frac{1}{2^{m-3}})$ separable. Furthermore, by the induction assumption, $\rho$ is separable. $\hfill\Box$\\

The separable balls for the multipartite quantum system have been studied in \cite{Gurvits2003-2}-\cite{Hildebrand2007}. The best one among them for the multiqubit system was given by Hildebrand \cite{Hildebrand2007}. It was claimed that, if a $m$-qubit unnormalized state $\rho$ satisfies
\begin{equation}
\label{Hil} ||\rho-I_{2^m}||_2\le \sqrt{\frac{54}{17}}\times 6^{-\frac{m}{2}}\times 2^m,
\end{equation}
then $\rho $ is separable.  Consider the state \[
\rho=I_2\otimes I_2\otimes \rho_3 \text{ with } \rho_3= \left( {{\begin{array}{*{20}c}
1& {\frac{1}{2}}  \\
 {\frac{1}{2}} &1\\
\end{array} }} \right).
\]
Some simple computations yield that $\rho$ is in the ball given by Corollary 3.5, but not in the ball given by (\ref{Hil}).

\section{Conclusions}
    In this paper, for the bipartite quantum system, we have put forward a  ball  of separable unnormalized states around the identity matrix, which is better than the ball in Frobenius norm. By using the special structure of the presented norm, the proposed ball has been used to give some separable conditions for pseudopure states and SPPT states, and, meanwhile, establish a new separable ball for the multipartite quantum system. The obtained simple separable conditions for pseudopure states and SPP states are the improvements or complements of the corresponding conditions in \cite{Gurvits2002}-\cite{Gurvits2005}, \cite{Chruscinski2008} and \cite{Bylicka2013}. An example of multiqubit state showed that there exist some states fall into the proposed separable ball for multipartite quantum system, but does not fall into the separable balls in \cite{Gurvits2003-2}-\cite{Hildebrand2007}.

There are some  problems that need to be addressed in the future. For instance, finding more applications of the separable ball given by Theorem 3.1 has important values. How to improve the radius of the presented separable ball for multipartite quantum system is an interesting problem.  A state  $\rho$ in $\mathbb{C}^{n_1}\otimes \mathbb{C}^{n_2}$ is said to be absolutely separable \cite{Kus2001} if $U^\dag \rho U$ is separable for any unitary matrix $U$ in $\mathbb{C}^{n_1}\otimes \mathbb{C}^{n_2}$. Clearly, any state in the ball with Frobenius norm is absolutely separable. A natural question is whether similar conclusion holds for the ball with the $(n_1,n_2)$-nested norm.
\section*{\bf Acknowledgments}
 This work is supported by the
NSFC (11105226,11326203,61403419), the Fundamental Research Funds for the
Central Universities (12CX04079A,24720122013), Research Award
Fund for outstanding young scientists of Shandong Province
(BS2012DX045). We are grateful to the referees and the editor for their invaluable suggestions to improve the quality of this paper. In particular, the referees pointed out some interesting problems that need to be further studied in the future.

\end{document}